\documentclass[a4paper,12pt]{article}
\usepackage{amssymb}

\begin{document}

\title{Machian Cosmological Solution in the Generalized Scalar-Tensor Theory of
Gravitation}
\author{A. Miyazaki \thanks{
Email: miyazaki@loyno.edu, miyazaki@nagasakipu.ac.jp} \vspace{3mm} \\
\textit{Department of Physics, Loyola University, New Orleans, LA 70118} \\
and \\
\textit{Faculty of Economics, Nagasaki Prefectural University} \\
\textit{Sasebo, Nagasaki 858-8580, Japan}}
\date{\vfill}
\maketitle

\begin{abstract}
The Machian cosmological solution satisfying $\phi =O(\rho /\omega )$ is
discussed for the homogeneous and isotropic universe with a perfect fluid
(with negative pressure) in the generalized scalar-tensor theory of
gravitation. We propose $\omega (\phi )=\eta /(\xi -2)$ for the coupling
function in the Machian point of view. The parameter $\xi $ varies in time
very slowly from $\xi =0$ to $\xi =2$ because of the physical evolution of
matter in the universe. When $\xi \rightarrow 2$, the coupling function
diverges to $-\infty $ and the scalar field $\phi $ converges to $G_{\infty
}^{-1}$. The present mass density is precisely predicted if the present time
of the universe is given. We obtain $\rho _{0}=1.6\times 10^{-29}\,g.cm^{-3}$
for $t_{0}=1.5\times 10^{10}\,yr$. The universe shows the slowly
decelerating expansion for the time-varying coupling function. \newline
\end{abstract}

\newpage

In the previous paper \cite{1)}, we discussed the Machian cosmological
solution satisfying $\phi =O(\rho /\omega )$ for the homogeneous and
isotropic universe with a perfect fluid in the Brans-Dicke theory \cite{2)}.
We found that the gravitational constant really approaches to constant when
the coefficient $\gamma $ of the equation of state goes to $-1/3$ (negative
pressure) for the closed model. If we assume the present mass density $\rho
_{0}$ is identical to the critical density $\rho _{c}$, taking $\left|
\omega \right| \sim 10^{3}$ \cite{3)} into account, we obtain the difference 
$\epsilon $ of the coefficient $\gamma $ from $-1/3$ ($\epsilon \equiv
3\gamma +1$)\ has a value of order $10^{-3}$ to support the present
gravitational constant. These parameters lead to the time-variation of the
gravitational constant $\left| \dot{G}/G\right| \sim 10^{-13}\,yr^{-1}$
which is compatible with the recent observational data \cite{4)}.

The closed model ($k=+1$) of this Machian solution is valid for the coupling
parameter $\omega $ satisfying \ $\omega /2+B<0$, where $B\equiv -3/(\xi
+2)(\xi -2)$ and $\xi \equiv 1-3\gamma $. The above parameter $\epsilon \sim
10^{-3}$ gives $B\sim 10^{3}$ which means $\omega <-10^{3}$ at present. The
expansion parameter $a(t)$ is explicitly a linear function of $t$, but its
coefficient depends on the parameter $\xi $. As the parameter $\xi $ goes to 
$2$ through the quasi-static process, the coefficient of the expansion
parameter increases very slowly for the constant coupling parameter $\omega $%
. Thus the universe exhibits the slowly accelerating expansion at present,
which is also compatible with the recent measurements \cite{5)} of the
distances to type Ia supernovae.

The above results strongly suggest that the coupling parameter $\omega $ of
the Brans-Dicke scalar field should vary in time as the universe expands. A
tentative conjecture is like $\omega (t)\sim B(\xi )$. The coupling
parameter $\omega $\ is almost constant and varies in time very slowly
through the quasi-static process of the parameter $\xi $. Therefore the
time-variation of the coupling parameter depends on the physical evolution
of matter in the universe.

In the present paper, we discuss the Machian cosmological solution and the
behavior of the coupling parameter $\omega (\phi )$ in the generalized
scalar-tensor theory of gravitation \cite{6)}-\cite{8)}. We consider the
homogeneous and isotropic universe filled with a perfect fluid with
pressure. We require a proper cosmological solution satisfies the postulate
that the scalar field $\phi $ should have the asymptotic form $\phi =O(\rho
/\omega )$ when the coupling parameter $\omega $ is large enough in order
that the scalar field converges to zero when $\left| \omega \right|
\rightarrow \infty $.

The action \cite{6)}-\cite{8)} for the generalized scalar-tensor theory of
gravitation is described in our sign conventions ($c=1$) as\newline
\begin{equation}
S=\int d^{4}x\sqrt{-g}[-\phi R+16\pi L_{m}-\frac{\omega (\phi )}{\phi }%
g^{\mu \nu }\phi _{,\,\mu }\phi _{,\,\nu }]\,,  \label{e1}
\end{equation}
where $R$ is the scalar curvature of the metric $g_{\mu \nu }$, $\phi (x)$
is the Brans-Dicke scalar field, $\omega (\phi )$ is an arbitrary coupling
function, and $L_{m}$ represents the Lagrangian for the matter fields. The
variation of Eq.(\ref{e1}) with respect to $g_{\mu \nu }$ and $\phi $ leads
to the field equations 
\begin{eqnarray}
R_{\mu \nu }-\frac{1}{2}Rg_{\mu \nu } &=&\frac{8\pi }{\phi }T_{\mu \nu }-%
\frac{\omega (\phi )}{\phi ^{2}}\left( \phi _{,\,\mu }\phi _{,\,\nu }-\frac{1%
}{2}g_{\mu \nu }\phi _{,\,\lambda }\phi ^{,\,\lambda }\right)  \nonumber \\
&&-\frac{1}{\phi }(\phi _{,\,\mu ;\,\nu }-g_{\mu \nu }\square \phi )\,,
\label{e2}
\end{eqnarray}
\begin{equation}
\square \phi =-\frac{1}{3+2\omega (\phi )}\left[ 8\pi T+\frac{d\omega (\phi )%
}{d\phi }\phi _{,\,\lambda }\phi ^{,\,\lambda }\right] \,,  \label{e3}
\end{equation}
where $T_{\mu \nu }$ is the energy-momentum tensor, $T$ is the contracted
energy-momentum tensor, and $\square $ denotes the generally-covariant
d'Alembertian $\square \phi \equiv \phi _{\;\;;\mu }^{,\mu }$. These field
equations satisfy the conservation law of the energy-momentum 
\begin{equation}
T_{;\nu }^{\mu \nu }=0\,.  \label{e4}
\end{equation}

The line element for the Friedmann-Robertson-Walker metric is 
\begin{equation}
ds^{2}=-dt^{2}+a^{2}(t)[d\chi ^{2}+\sigma ^{2}(\chi )(d\theta ^{2}+\sin
^{2}\theta d\varphi ^{2})]\,,  \label{e5}
\end{equation}
where 
\begin{equation}
\sigma (\chi )\equiv \left\{ 
\begin{array}{l}
\sin \chi \;\;\;\;\;for\;k=+1\;(closed\;space) \\ 
\chi \;\;\;\;\;\;\;\;for\;k=0\;\ \;(flat\;space) \\ 
\sinh \chi \;\;\;for\;k=-1\;\ (open\;space)\,.
\end{array}
\right.  \label{e6}
\end{equation}
The energy-momentum tensor for the perfect fluid with pressure $p$ is given
as 
\begin{equation}
T_{\mu \nu }=-pg_{\mu \nu }-(\rho +p)u_{\mu }u_{\nu }\,,  \label{e7}
\end{equation}
where $\rho $\ is the mass density in comoving coordinates and $u^{\mu }$ is
the four velocity $dx^{\mu }/d\tau $ ($\tau $ is the proper time). The
nonvanishing components are $T_{00}=-\rho $, $T_{i\,i}=-pg_{i\,i}$ ($i\neq 0$%
), and its trace is $T=\rho -3p$ for the homogeneous and isotropic universe.

The energy conservation Eq.(\ref{e4}) leads to the equation of continuity 
\begin{equation}
\dot{\rho}+3\frac{\dot{a}}{a}\left( \rho +p\right) =0\,.  \label{e8}
\end{equation}
Let us suppose the equation of state for the perfect fluid 
\begin{equation}
p(t)=\gamma \rho (t)\,,  \label{e9}
\end{equation}
where $-1\leqq \gamma \leqq 1/3$. Now we are rather interested in the
''negative'' pressure. Taking the equation of state into account, we can
integrate the equation of continuity Eq.(\ref{e8}) and get 
\begin{equation}
\rho (t)a^{n}(t)=const\,,  \label{e10}
\end{equation}
where $n=3(\gamma +1)$, which has $n=4$, $n=3$, and $n=2$ for the radiation
era, the matter-dominated era, and the negative pressure $\gamma =-1/3$ era
respectively.

The independent field equations which we need solve simultaneously are 
\begin{equation}
\frac{3}{a^{2}}\left( \dot{a}^{2}+k\right) =\frac{\omega (\phi )}{2}\left( 
\frac{\dot{\phi}}{\phi }\right) ^{2}+\frac{\ddot{\phi}}{\phi }+\frac{8\pi
\rho }{\phi }-\frac{8\pi \left( \rho -3p\right) }{3+2\omega (\phi )}\frac{1}{%
\phi }\,,  \label{e11}
\end{equation}
and 
\begin{equation}
\ddot{\phi}+3\frac{\dot{a}}{a}\dot{\phi}=\frac{1}{3+2\omega (\phi )}\left[
8\pi \left( \rho -3p\right) -\frac{d\omega (\phi )}{d\phi }\dot{\phi}^{2}%
\right] \,.  \label{e12}
\end{equation}

According to the postulate $\phi =O(\rho /\omega )$, we expect a solution
described as 
\begin{equation}
\phi (t)=\frac{8\pi }{3+2\omega (\phi )}\Phi (t)\,,  \label{e13}
\end{equation}
where $\Phi (t)$ is an unknown function of $t$\ and may generally include $%
\omega $ as the following form 
\begin{equation}
\Phi (t)=\Phi _{0}(t)+O(1/\omega )\,.  \label{e14}
\end{equation}
In the Brans-Dicke theory, we know that $\Phi (t)$ should not include $%
\omega $\ for the perfect fluid model \cite{9)}, \cite{1)}.

The most important thing is to determine an arbitrary coupling function $%
\omega (\phi )$. There are many literatures, for example \cite{10)}-\cite
{15)}. We seek some possibilities in the Machian point of view. First, if
the coupling function $\omega (\phi )$ obeys 
\begin{equation}
\frac{d\omega (\phi )}{d\phi }\dot{\phi}^{2}=8\pi \eta \rho (t)\,,
\label{e15}
\end{equation}
we can adopt the similar procedure \cite{9)} for the Machian cosmological
solution in the Brans-Dicke theory. The coefficient $\eta $ is an arbitrary
constant.

Let us assume that the following relation is satisfied approximately enough
because the coupling function $\omega (\phi )$ is almost constant or large
enough: 
\begin{equation}
\dot{\phi}(t)=\frac{8\pi }{3+2\omega (\phi )}\dot{\Phi}(t)\,.  \label{e16}
\end{equation}
We need check later whether this simplification is valid or not for the
obtained Machian solution. Taking Eq.(\ref{e16}) into account, we get from
Eqs.(\ref{e12}) and (\ref{e15}) 
\begin{equation}
\ddot{\Phi}+3\frac{\dot{a}}{a}\dot{\Phi}=\left( \xi +\eta \right) \rho \,,
\label{e17}
\end{equation}
where $\xi =1-3\gamma $ or$\ \xi =4-n$. The function $\Phi (t)$ should not
include $\omega $ similarly in the Brans-Dicke theory for the Machian
solution because the right-hand side of Eq.(\ref{e15}) does not depend on $%
\omega $. Thus the ratio $\dot{a}/a$ should not also include $\omega $, and
so we find for the expansion parameter\ 
\begin{equation}
a(t)\equiv A(\omega )\alpha (t)\,,  \label{e18}
\end{equation}
where $A$ and $\alpha $\ are arbitrary functions of only $\omega $\ and $t$\
respectively.

Taking Eqs.(\ref{e13}), (\ref{e16}), and (\ref{e18}), we obtain from Eq.(\ref
{e11}) after elimination of $\ddot{\phi}$ by Eq.(\ref{e12}) 
\begin{equation}
\frac{\omega }{2}\left[ \left( \frac{\dot{\Phi}}{\Phi }\right) ^{2}+\frac{%
4\rho }{\Phi }\right] -\frac{3k}{A^{2}(\omega )\alpha ^{2}}=3\left( \frac{%
\dot{\alpha}}{\alpha }\right) ^{2}+3\left( \frac{\dot{\alpha}}{\alpha }%
\right) \left( \frac{\dot{\Phi}}{\Phi }\right) -\frac{\left( 3+\eta \right)
\rho }{\Phi }\,.  \label{e19}
\end{equation}
We are not interested in the flat space ($k=0$) here. For the closed and the
open spaces ($k=\pm 1$), if we require that Eq.(\ref{e19}) is identically
satisfied for all arbitrary values of $\omega $, we find that\ the
coefficient $A(\omega )$ must have the following form 
\begin{equation}
\frac{3}{A^{2}(\omega )}=\left| \frac{\omega (\phi )}{2}+B\right| \,,
\label{e20}
\end{equation}
where $B$\ is a constant with no dependence of $\omega $. Thus we get\newline
\begin{equation}
\frac{\omega }{2}\left[ \left( \frac{\dot{\Phi}}{\Phi }\right) ^{2}+\frac{%
4\rho }{\Phi }-k\,j\frac{1}{\alpha ^{2}}\right] =3\left( \frac{\dot{\alpha}}{%
\alpha }\right) ^{2}+3\left( \frac{\dot{\alpha}}{\alpha }\right) \left( 
\frac{\dot{\Phi}}{\Phi }\right) -\frac{\left( 3+\eta \right) \rho }{\Phi }%
+k\,j\frac{B}{\alpha ^{2}}\,,  \label{e21}
\end{equation}
where we introduce a notation $j=-1$ for $\omega /2+B<0$ and $j=+1$ for $%
\omega /2+B>0$. Again, by requiring that Eq.(\ref{e21}) is identically
satisfied for all $\omega $, we obtain two identities, 
\begin{equation}
\left( \frac{\dot{\Phi}}{\Phi }\right) ^{2}+\frac{4\rho }{\Phi }\equiv k\,j%
\frac{1}{\alpha ^{2}}\,,  \label{e22}
\end{equation}
and 
\begin{equation}
3\left( \frac{\dot{\alpha}}{\alpha }\right) ^{2}+3\left( \frac{\dot{\alpha}}{%
\alpha }\right) \left( \frac{\dot{\Phi}}{\Phi }\right) -\frac{\left( 3+\eta
\right) \rho }{\Phi }\equiv -k\,j\frac{B}{\alpha ^{2}}\,.  \label{e23}
\end{equation}

We are convinced of the existence of the following type of solutions on the
analogy of the previous paper \cite{1)}: 
\begin{equation}
\Phi (t)=\zeta \rho (t)t^{2}\,,  \label{e24}
\end{equation}
\begin{equation}
\alpha (t)=bt\,,  \label{e25}
\end{equation}
where coefficients $\zeta $\ and $b$\ are constants respectively, and in
fact we find 
\begin{equation}
\zeta =\frac{\xi +\eta }{\xi (\xi -2)}\,,  \label{e26}
\end{equation}
\begin{equation}
b^{2}=\left\{ 
\begin{array}{l}
-\frac{\xi +\eta }{(\xi -2)\left[ \xi ^{2}+(\eta +2)\xi -2\eta \right] }%
\,,\;\;for\;k\,j=-1\;and\;0\leqq \xi <2 \\ 
\frac{\xi +\eta }{(\xi -2)\left[ \xi ^{2}+(\eta +2)\xi -2\eta \right] }%
\,,\;\;for\;k\,j=+1\;and\;2<\xi \leqq 4\,,\,
\end{array}
\right.  \label{e27}
\end{equation}
and 
\begin{equation}
B=\frac{\eta \xi ^{2}-(5\eta +3)\xi +3\eta }{(\xi -2)\left[ \xi ^{2}+(\eta
+2)\xi -2\eta \right] }  \label{e28}
\end{equation}
from Eqs.(\ref{e17}), (\ref{e22}), and (\ref{e23}) successively.

For the closed space ($k=+1$), we require $\eta >0$ for $\zeta <0$ when $%
0\leqq \xi <2$ to give the attractive gravitational force ($G>0$). On the
other hand we require $\eta <0$, for the coefficient $b$ is real when $%
0\leqq \xi <2$. So we encounter a contradiction. Let us adopt the other
alternative that the coefficient $\eta \ $is a linear function of $\xi $ ($%
\eta =\eta _{1}\xi +\eta _{0}$). We put $\eta _{0}=0$ for the finite $\zeta $
when $\xi =0$. Thus we replace Eq.(\ref{e15}) to 
\begin{equation}
\frac{d\omega (\phi )}{d\phi }\dot{\phi}^{2}=8\pi \eta _{1}\xi \rho (t)=8\pi
\eta _{1}\left( \rho -3p\right) \,.  \label{e29}
\end{equation}
After similar discussions, we obtain as a solution 
\begin{equation}
\zeta =\left( \eta _{1}+1\right) /(\xi -2)\,,  \label{e30}
\end{equation}
\begin{equation}
b^{2}=\left\{ 
\begin{array}{l}
-1/(\xi -2)(\xi -\xi _{1})\,,\;\;for\;k\,j=-1\;and\;0\leqq \xi <2 \\ 
1/(\xi -2)(\xi -\xi _{1})\,,\;\;for\;k\,j=+1\;and\;2<\xi \leqq 4\,,
\end{array}
\right.  \label{e31}
\end{equation}
and 
\begin{equation}
B=[\eta _{1}(\xi ^{2}-5\xi +3)-3]/(\xi -2)\left[ (\eta _{1}+1)\xi -2(\eta
_{1}-1)\right] \,,  \label{e32}
\end{equation}
where $\xi _{1}\equiv 2(\eta _{1}-1)/\left( \eta _{1}+1\right) $.

We require $-1<\eta _{1}<1$ for $\zeta <0$ and the real $b$ when $0\leqq \xi
<2$ ($k=+1$). By analyzing the time-dependences of the solution, we get from
Eq.(\ref{e29})\newline
\begin{equation}
\frac{d\omega }{d\phi }=\frac{\eta _{1}\xi (3+2\omega )}{\left( \eta
_{1}+1\right) (\xi -2)}\frac{1}{\phi }\,.  \label{e33}
\end{equation}
After integration, we obtain the coupling function 
\begin{equation}
\left| 3+2\omega (\phi )\right| =C\phi (t)^{\left[ 2\eta _{1}\xi /\left(
\eta _{1}+1\right) (\xi -2)\right] }\,,  \label{e34}
\end{equation}
where $C$ is an integral constant. However, we realize that this coupling
function $\omega (\phi )$ is not consistent with Eq.(\ref{e16}) because $%
\omega (\phi )$ varies in time too rapidly.

As the next alternative, we suppose 
\begin{equation}
\frac{d\omega (\phi )}{d\phi }\dot{\phi}^{2}=\frac{8\pi \eta \rho (t)}{%
3+2\omega (\phi )}  \label{e35}
\end{equation}
to suppress the rapid time-variation of $\omega (\phi )$. In this case, the
right-hand side of Eq.(\ref{e17}) includes the coupling function, and so the
function $\Phi (t)$ may also include $\omega (\phi )$ as Eq.(\ref{e14}).
From now on, we restrict our calculations to the first order in $%
1/(3+2\omega )$. At least for the present time ($\left| \omega \right|
\gtrsim 10^{3}$), this restriction would give a good approximation. For the
first order in $1/(3+2\omega )$, the function $\Phi (t)$ does not include $%
\omega (\phi )$. Thus we obtain the Machian solution for the first order in $%
1/(3+2\omega )$ again, 
\begin{equation}
\zeta =1/(\xi -2)\,,  \label{e36}
\end{equation}
\begin{equation}
b=\left\{ 
\begin{array}{l}
(4-\xi ^{2})^{-1/2}\,,\;\;for\;k\,j=-1\;and\;0\leqq \xi <2 \\ 
(\xi ^{2}-4)^{-1/2}\,,\;\;for\;k\,j=+1\;and\;2<\xi \leqq 4\,,
\end{array}
\right.  \label{e37}
\end{equation}
and 
\begin{equation}
B=-3/(\xi -2)(\xi +2)\,.  \label{e38}
\end{equation}

In Eqs.(\ref{e33}) or (\ref{e34}), the scalar field $\phi $ includes the
coupling function $\omega (\phi )$ itself. This expression causes
ambiguities, and so let us define another function of $\Phi $%
\begin{equation}
\omega (\phi )=\varpi (\Phi )\,.  \label{e39}
\end{equation}
Taking this expression into account, we estimate Eq.(\ref{e35}) by the
Machian solution for the first order in $1/(3+2\omega )$ and get 
\begin{equation}
\frac{d\varpi }{d\Phi }=\frac{\eta }{\xi -2}\frac{1}{\Phi }\,,  \label{e40}
\end{equation}
which leads to 
\begin{equation}
\varpi (\Phi )=\frac{\eta }{\xi -2}\ln \left| \Phi (t)\right| +W\,,
\label{e41}
\end{equation}
where $W$ is an integral constant.

When $\xi \rightarrow 2$, the time-dependence of $\rho (t)$ goes to $t^{-2}$
and so $\rho (t)t^{2}\rightarrow const$. However, the coefficient $\zeta $
diverges to the minus infinity as $\xi \rightarrow 2$ in the range of $%
0\leqq \xi <2$. So the function $\Phi (t)=\zeta \rho (t)t^{2}$ also diverges
to the minus infinity very slowly as $\xi \rightarrow 2$ in the same range.
Thus, if $\eta >0$, the coupling function $\varpi (\Phi )$ diverges to the
minus infinity when $\xi \rightarrow 2$ in this range. On the other hand,
when $t\rightarrow 0$ and $\xi \rightarrow 0$, $\Phi (t)$ and $\varpi (\Phi
) $ also diverge to the minus infinity. The time-derivative of the scalar
field $\phi (t)$ is described in the present solution as\newline
\begin{equation}
\dot{\phi}(t)=\frac{8\pi }{3+2\varpi (\Phi )}\dot{\Phi}(t)\left[ 1-\frac{16}{%
3+2\varpi (\Phi )}\frac{\eta }{\xi -2}\right] \,,
\end{equation}
and the second term of the middle bracket ($\sim 1/\ln \left| \Phi
(t)\right| $) goes to zero when $\xi \rightarrow 0$ and $\xi \rightarrow 2$.
Therefore the assumption Eq.(\ref{e16}) is consistent in our Machian
solution near $\xi =0$ and $\xi =2$. The coupling function Eq.(\ref{e41})
satisfies the constraint 
\begin{equation}
\frac{\varpi (\Phi )}{2}+B<0\,,  \label{e43}
\end{equation}
which leads to 
\begin{equation}
\eta \ln \left| \Phi (t)\right| +(\xi -2)W>6/(\xi +2)\,,  \label{e44}
\end{equation}
if $\eta >0$ in the range of $0\leqq \xi <2$ (at least near $\xi =0$ and $%
\xi =2$, or for the appropriate $W$). The behavior of the scalar field
becomes 
\begin{equation}
\phi (t)\sim (\xi -2)\frac{\Phi (t)}{\ln \left| \Phi (t)\right| }=\frac{\rho
(t)t^{2}}{\ln \left| \zeta \rho (t)t^{2}\right| }\rightarrow
const\rightarrow 0  \label{e45}
\end{equation}
when $\xi \rightarrow 2$. The scalar field is approaching to $const$ as the
parameter $\xi \rightarrow 2$ and its value gradually converges to zero.

The expansion parameter is expressed as 
\begin{equation}
a(t)\equiv A(\omega )bt=\left[ \frac{6}{\varpi (\Phi )(\xi +2)(\xi -2)-6}%
\right] ^{1/2}t  \label{e46}
\end{equation}
for the closed space ($k=+1$) in the range of $0\leqq \xi <2$. Taking $%
\varpi (\Phi )\rightarrow -\infty $ when $t\rightarrow 0$ into account, we
get $a(t)\approx 0$ near $\xi =0$. By replacing $\eta $ to $\eta _{1}\xi $,
we might be able to improve the situation, that is,\newline
\begin{equation}
\frac{d\omega (\phi )}{d\phi }\dot{\phi}^{2}=\frac{8\pi \eta _{1}\left( \rho
-3p\right) }{3+2\omega (\phi )}\,,  \label{e47}
\end{equation}
and thus 
\begin{equation}
\varpi (\Phi )=\frac{\eta _{1}\xi }{\xi -2}\ln \left| \Phi (t)\right| +W\,.
\label{e48}
\end{equation}

When we consider the case near $\xi =2$, we encounter a severe trouble: the
coefficient of $a(t)$\ in Eq.(\ref{e46}) converges to zero as $\Phi
(t)\rightarrow -\infty $ when $t\rightarrow \infty $. To satisfy the
constraint Eq.(\ref{e43}), the coupling function $\varpi (\Phi )$ must
decrease faster than $B\sim 1/(\xi -2)$ when $\xi \rightarrow 2$. On the
other hand, the coupling function $\varpi (\Phi )$ must decrease slower than 
$1/(\xi -2)$ when $\xi \rightarrow 2$ in order to give the slowly
accelerating expansion. After all, there are no alternatives but 
\begin{equation}
\varpi (\Phi )=\frac{\eta }{\xi -2}\,,  \label{e49}
\end{equation}
(which would give the almost linear expansion,) for the Machian cosmological
solution. This means 
\begin{equation}
\frac{d\varpi }{d\Phi }=0\,.  \label{e50}
\end{equation}
The coupling function $\varpi (\Phi )$ does not depend on $\Phi $. Therefore
the solutions (\ref{e24}) and (\ref{e25}) with Eqs.(\ref{e36}), (\ref{e37}),
and (\ref{e38}) are exact for all coupling parameter $\omega $. The
constraint Eq.(\ref{e43}) for all $\xi $ ($0\leqq \xi <2$) requires the
condition $\eta >3$, which gives $\omega <-3/2$ when $\xi =0$ and avoids the
singularity. For this coefficient $\eta >3$, the expansion parameter $a(t)$
gives the extremely slowly decelerating expansion except the early stage
with $\xi =0$. We obtain $a(t)=t$ for $\xi \rightarrow 2$ if $\eta =3$.

The scalar field $\phi $ for the coupling function $\varpi (\Phi )=\eta
/(\xi -2)$ is given as the following and we get when $\xi \rightarrow 2$%
\begin{equation}
\phi (t)=\frac{8\pi \rho (t)t^{2}}{3(\xi -2)+2\eta }\cong \frac{4\pi \rho
(t)t^{2}}{\eta }\rightarrow const\,,  \label{e51}
\end{equation}
which converges to a definite and finite constant in the limit. If we assume 
$\eta =3$, taking $t_{0}=1.0\times 10^{10}\,yr$ and the present
gravitational constant $G_{0}=6.67\times 10^{-8}\,dyn.cm^{2}.g^{-1}$ into
account, we can estimate the present mass density $\rho _{0}=3.5\times
10^{-29}\,g.cm^{-3}$. If we adopt $t_{0}=1.5\times 10^{10}\,yr$, we obtain $%
\rho _{0}=1.6\times 10^{-29}\,g.cm^{-3}$, which is very near to the critical
density $\rho _{c}\sim 10^{-29}\,g.cm^{-3}$. As the parameter $\xi
\rightarrow 2$, the coupling function $\varpi (\Phi )$ diverges to the minus
infinity and the gravitational constant approaches dynamically to the
constant $G_{\infty }$.

It should be remarked that the time-variation of the coupling function $%
\varpi (\Phi )=\eta /(\xi -2)$ is derived from that of the parameter $\xi $.
However, we have not known the details of the physical evolution of matter
in the universe yet. Probably, our universe started (classically) from the
Big Bang with $\xi =0$ (the radiation era), via the matter-dominated era ($%
\xi =1$), and now must be approaching to the negative pressure era (the 
\emph{quintessence} era, $\gamma =-1/3$, $\xi =2$). According to the recent
measurements \cite{3)} of the coupling parameter ($\left| \omega \right|
\sim 10^{3}$), we get $\epsilon \equiv 2-\xi \sim 10^{-3}$ from Eq.(\ref{e49}%
). Over $10^{10}\,yr$, the state of the universe has been varying extremely
slowly from $\xi =0$ to $\epsilon \sim 10^{-3}$. This situation would solve
the problems existing in the solar system and the evolution of life in the
time-varying gravitational constant. At $t\sim 5\times 10^{9}\,yr$, the
parameter $\epsilon $ had already been small enough and the gravitational
constant did not differ much from the present value.

There is a discontinuity at $\xi =2$ for the Machian cosmological solution,
and so it is not reasonable that the universe continues to evolve beyond $%
\xi =2$ to the state $\xi =4$. Our universe must be approaching to the state 
$\xi =2$ for ever ($t\rightarrow +\infty $). We require that the universe is
closed ($k=+1$ ) for this range of $\xi $ ($0\leqq \xi <2$) to give the
attractive gravitational force. As the parameter $\xi $ varies in time
extremely slowly (the quasi-static process) in this range, we may regard
that the parameter $\xi $ is constant when we execute the derivative with
respect to $t$.

It is believed that the generalized scalar-tensor theory of gravitation
reduces to general relativity with the same energy-momentum tensor when $%
\left| \omega \right| \rightarrow +\infty $ and $\omega ^{-3}d\omega /d\phi
\rightarrow 0$, but this is not generally true. The present Machian
cosmological solution has no ordinary correspondences in general relativity
even when $\left| \omega \right| \rightarrow +\infty $ and $\omega
^{-3}d\omega /d\phi \rightarrow 0$. The scalar field of Machian solutions
satisfies the asymptotic form $\phi =O(\rho /\omega )$ and converges
generally to zero when $\left| \omega \right| \rightarrow +\infty $ or $\rho
\rightarrow 0$. It should be noted that the scalar field $\phi (t)$
converges to the finite $G_{\infty }^{-1}$ in the present Machian solution
when $\omega \rightarrow -\infty $ and $\rho \rightarrow 0$ as $t\rightarrow
+\infty $, though it satisfies $\phi =O(\rho /\omega )$. The parameter $%
\omega $ is no more arbitrary but is determined by the evolution of the
universe in the generalized scalar-tensor theory of gravitation.

When we deal exactly with the time-varying coupling function $\varpi (\Phi
)=\eta /(\xi -2)$, the expansion parameter $a(t)$ does not show the slowly
accelerating expansion but rather the extremely slowly decelerating
expansion (almost the linear expansion). If this is not compatible with the
recent observations, we might need to introduce furthermore the cosmological
constant to the present Machian cosmological model.

We have no criterions to determine the coupling function $\omega (\phi )$ in
the generalized scalar-tensor theory of gravitation. We discussed some
examples here. They seem to suggest that it is not an arbitrary function of $%
\phi $ but another scalar field which is derived from matter itself by
another field equation. The scalar field $\phi $ was introduced to the
Brans-Dicke theory through the relation $GM/c^{2}R\sim 1$. By the similar
analogy, the relation $\omega \sim 1/(\xi -2)$ seems to require the
existence of another unknown scalar field connected with matter.

The next straightforward problem is to determine the time-variation and the
range of the parameter $\xi (t)$. We need investigate the evolution of a
scalar field as the dark matter in the universe.\newline
\newline
\textbf{Acknowledgment}

The author is grateful to Professor Carl Brans for helpful discussions and
his hospitality at Loyola University (New Orleans) where this work was done.
He would also like to thank the Nagasaki Prefectural Government for
financial support.

\end{document}